# Impedance-matched low-pass stripline filters

D F Santavicca and D E Prober
Department of Applied Physics, Yale University, New Haven, CT 06520-8284

E-mail: daniel.santavicca@yale.edu, daniel.prober@yale.edu

**Abstract.** We have constructed several impedance-matched low-pass filters using a stripline geometry with a dissipative dielectric. The filters are compact, simple to construct, and operate in cryogenic environments. The dissipative dielectric consists of magnetically-loaded silicone or epoxy, which are commercially available under the trade name Eccosorb. For a stripline length of 32 mm, the filters have a passband that extends from dc to a 3 dB bandwidth between 0.3 and 0.8 GHz. The 3 dB bandwidth can be adjusted beyond this range by changing the filter length. An extremely broad stopband at higher frequencies, with attenuation exceeding 100 dB, is achieved along with a return loss greater than 10 dB measured up to 40 GHz. This combination of high attenuation and low reflected power across a broad stopband ensures that spurious or unwanted signals outside the passband do not reach or return to the device under test. This type of filter has applications in microwave frequency measurements of sensitive non-linear devices such as superconducting heterodyne mixers, quantum tunneling devices, and quantum computing elements.

## 1. Introduction

The magnetically lossy transmission line as a low-pass filter has been known for many years [1,2]. This type of filter is often used to achieve very broad high frequency attenuation combined with negligible dc resistance, e.g. for filtering the leads of power supplies. We have developed a magnetically lossy transmission line filter for use in microwave frequency measurements of strongly non-linear quantum devices. These devices are extremely sensitive to high frequency noise and hence require careful cryogenic filtering [3]. The termination impedance seen by the device can also have a significant effect on its performance. Termination by an incorrect impedance may have a detrimental effect that is difficult to model or predict. An ideal filter will not only attenuate strongly out to very high frequency, it will also maintain a matched input impedance. Our filter was developed with an emphasis on maintaining a large return loss, and hence an impedance close to 50 Ω, over many decades in frequency. We characterize these filters both at room temperature and at cryogenic temperatures.

One application of such a filter is in the microwave characterization of heterodyne mixers, including superconducting tunnel junction and bolometric mixers [4]. As a non-linear device, the heterodyne mixer responds at multiple frequencies in addition to the input frequencies. To avoid spurious effects, the local oscillator power reflected by the mixer and also the higher harmonics generated and emitted by the mixer need to be properly terminated at the intermediate frequency port. The impedance-matched low-pass filter ensures a low level of reflected power out to very high frequency while efficiently transmitting the intermediate frequency output, which is in the filter's passband. Another potential application is filtering the lines used to manipulate quantum bits with fast pulse sequences, as well as the high-frequency readout lines, provided that the filter bandwidth is sufficiently large.

Conventional filters made from reactive lumped element components have a stopband that is limited by parasitics. They are also not designed to be impedance-matched in their stopband. When measuring highly sensitive cryogenic devices, it is often important to attenuate room temperature thermal (blackbody) noise. For coupling via a 1D transmission line such as a coaxial cable, the noise power $P_{th}$ per unit bandwidth B is given by the 1D Planck spectrum,

$$\frac{P_{th}}{B} = \frac{hf}{e^{hf/k_B T} - 1}. \tag{1}$$

At T = 300 K, this extends in frequency up to ~10 THz. Cable losses effectively attenuate the highest frequencies, but a filter with a stopband extending to >100 GHz is desirable. It is difficult to achieve effective filtering at frequencies >> 10 GHz with lumped element components due to their parasitic reactance. To filter up to ~100 GHz, an alternative approach is required.

A much broader stopband can be achieved by using spatially distributed circuit elements instead of lumped-element components. Several types of non-impedance-matched distributed filters have been reported, primarily for filtering the biasing lines of single-electron and quantum computing circuits. The most widely used is the metal powder filter, which consists of a long wire embedded in a mixture of metal powder and epoxy [5-8]. This low-pass filter utilizes the capacitance between the wire and the powder as well as eddy-current dissipation in the powder and has a typical 3 dB bandwidth ~1 MHz. (The 3 dB bandwidth is the frequency at which the attenuation is 3 dB, or approximately ½ in linear units.) Microfabricated versions of such non-impedance-matched distributed filters have also been reported [3,9,10].

To achieve impedance matching, a distributed filter must be constructed in a transmission line geometry with frequency-dependent dissipation in the conductor or the dielectric. Previous work described a filter based on the resistive coaxial cable Thermocoax [11] and a metal powder filter in a coaxial geometry [12]. We describe a filter that uses a magnetically-loaded dielectric in a stripline geometry. This filter exploits a combination of magnetic and dielectric dissipation to achieve significantly greater attenuation per unit length in the stopband than previous impedance-matched filters. The stripline filter also demonstrates a return loss greater than 10 dB (less than 10% reflected power) out to 40 GHz. The performance of these three types of impedance-matched low-pass filters is compared in table 1.

**Table 1.** Comparison of lossy transmission line filters.

| Filter | DC resistance per unit length ($\Omega$/m) | Attenuation per unit length at 10 GHz (dB/m) | Return loss at 10 GHz (dB) |
|---|---|---|---|
| Thermocoax cable [11] | ~50 | ~175 | ~6 |
| Coaxial metal powder filter [12] | < 1 | ~590 | >10 |
| Stripline filter (this work) | < 1 | >3000 | >10 |

**2. Filter Design**

A stripline geometry was chosen because it is easier to construct than a coaxial geometry and, unlike a microstrip, the field lines are entirely inside the dielectric, maximizing the dissipation. The stripline supports a TEM mode at low frequency, with higher order modes appearing when the enclosure begins to act as a waveguide.

To determine the attenuation constant, we consider a TEM mode in a transmission line with a complex permittivity and permeability. The permittivity can be expressed as $\varepsilon = \varepsilon' - j\varepsilon'' = |\varepsilon| e^{j\delta_\varepsilon}$, where $\tan \delta_\varepsilon = \varepsilon''/\varepsilon'$ is the dielectric loss tangent. Similarly, the permeability can be expressed as $\mu = \mu' - j\mu'' = |\mu| e^{j\delta_\mu}$, where $\tan \delta_\mu = \mu''/\mu'$ is the magnetic loss tangent. The complex propagation constant is $\gamma = j\omega\sqrt{\varepsilon\mu} = \alpha + j\beta$, where ω is the angular frequency, α is the attenuation constant, and β is the phase constant. For propagation in the x direction, the electric field magnitude is $E(x) = E(0) e^{-\gamma x}$. We can express the attenuation constant $\alpha = \text{Re}[\gamma]$, which can be written using the above relations as

$$\alpha = (1.48 \times 10^{-8}) f \sqrt{\frac{\mu'\varepsilon'}{\mu_0 \varepsilon_0}} \left\{ \sqrt{\left[1 + \left(\frac{\varepsilon''}{\varepsilon'}\right)^2\right]\left[1 + \left(\frac{\mu''}{\mu'}\right)^2\right]} - 1 + \left(\frac{\varepsilon''}{\varepsilon'}\right)\left(\frac{\mu''}{\mu'}\right) \right\}^{1/2} \quad (2)$$

in units of Np/m, where f is the frequency and $\mu_0$ and $\varepsilon_0$ are the free space permeability and permittivity, respectively. The transmitted power is thus $P(x) = P(0) e^{-2\alpha x}$. The attenuation due to the dielectric material depends on the length but is otherwise independent of the transmission line geometry and is only a function of the dielectric material properties. Magnetically-loaded dielectrics provide very high attenuation because they typically have a magnetic loss tangent that increases with frequency up to ~1 GHz combined with a large permittivity [13].

The magnetically-loaded dielectrics used in our filters are from the Eccosorb line of microwave absorbing materials from Emerson & Cuming Microwave Products [14]. They come in carbon- and magnetically-loaded varieties in different thicknesses and with several dielectric embedding materials. We used the FGM-40, GDS, and MCS materials, which are magnetically-loaded silicone sheets, as well as the CR-124 material, which is a magnetically-loaded epoxy. The magnetic loading material is carbonyl iron or ferrite powder, or a combination of the two.

A rectangular enclosure, machined in a block of copper, defines the outer dimensions of the stripline. The enclosure is filled with the Eccosorb material. Two SMA receptacles with an extended dielectric and pin contact [15] are mounted on each end of the copper block such that the pins extend into the center of the enclosure. Connecting the pins is a strip cut from a sheet of 0.15 mm thick Cu foil. The ends of the strip are soldered to each pin contact. The silicone material has the advantage of allowing the strip to be modified after construction of the filter, as the strip is simply sandwiched between two layers of material. An image of two completed filters made with the silicone Eccosorb material is presented in figure 1.

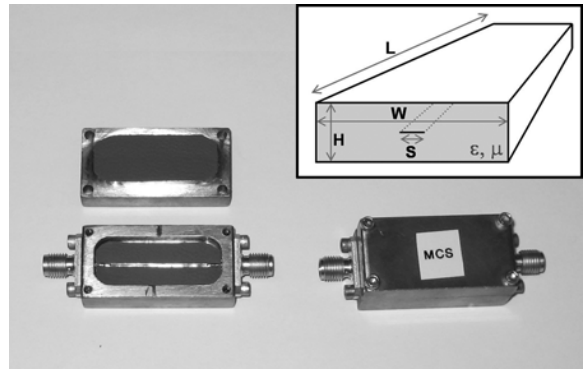

**Figure 1.** Photograph of two completed filters, one open and one closed. Inset: sketch of the stripline geometry, shown in a cross-sectional cut viewed from the end.

The dimensions of each stripline filter, as illustrated in figure 1, are summarized in table 2. The center conductor width S was chosen by testing strips of different widths and selecting the one that gave the greatest average return loss in the measurement range 50 MHz – 40 GHz after a SOLT (short, open, load, thru) calibration at the coaxial connector reference plane. Introducing a slight taper on each end of the center conductor further improved the reflection coefficient. We used a linear taper for simplicity and found that the greatest return loss was obtained with a taper that transitions from the width of the pin to the width of the center conductor in a length of 2-3 mm. The stripline geometry assumes that the side walls do not significantly alter the field lines. This condition is valid if the width of the stripline enclosure W is much greater than the width of the center conductor S. In our design, the enclosure width is always more than ten times the width of the center conductor.

**Table 2.** Filter dimensions.

| Dielectric Material | S (mm) | H (mm) | W (mm) | L (mm) |
|---|---|---|---|---|
| CR-124 | 1.4 | 4.6 | 18 | 32 |
| FGM-40 | 1.1 | 2.0 | 13 | 32 |
| GDS | 0.8 | 1.5 | 13 | 32 |
| MCS | 1.1 | 2.0 | 13 | 32 |

**3. Filter Characterization**

The attenuation and return loss of the stripline filters made from the four different Eccosorb materials were measured at room temperature, 77 K, and 4.2 K using an HP 8722D 50 MHz – 40 GHz network analyzer. The attenuation is defined as *-10 log ($P_{trans}/P_{inc}$)*, where $P_{trans}$ is the power measured at the output of the filter and $P_{inc}$ is the power incident on the input of the filter. The return loss is defined as *-10 log ($P_{ref}/P_{inc}$)*, where $P_{ref}$ is the power that is reflected off the filter input. For testing at 77 K and 4.2 K, the filter was immersed in liquid nitrogen and liquid helium, respectively. The filter performance was unchanged after multiple thermal cycles. The attenuation of each filter is presented in figure 2. The 3 dB bandwidth ($f_{3dB}$) increases as the temperature is lowered for all four filters. This is likely due to a decrease in the permittivity with decreasing temperature.

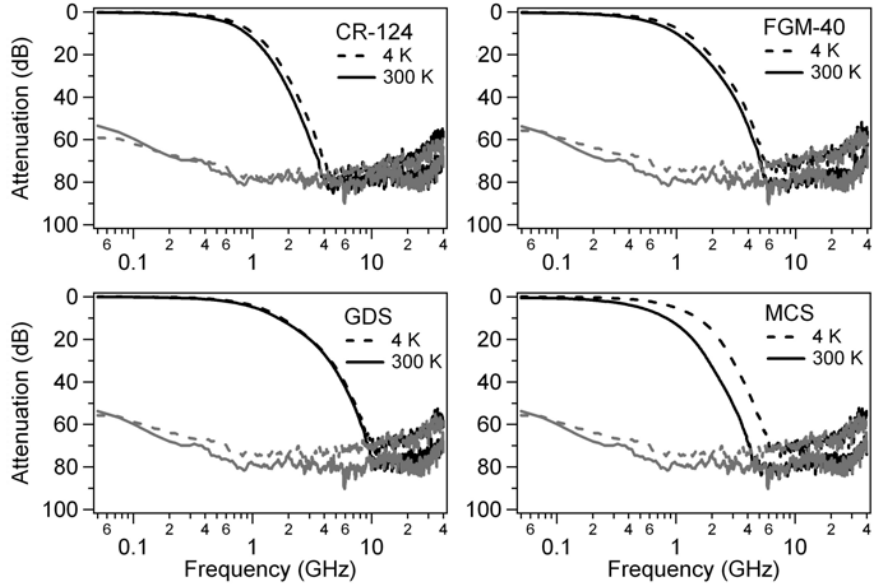

**Figure 2.** Measured attenuation of the CR-124, FGM-40, GDS, and MCS stripline filters at room temperature (solid line) and 4.2 K (dashed line). The instrument noise floor is indicated in grey. (The noise floor is higher at 4.2 K because of the loss of the extra cables used to measure at low temperature.)

The signal can no longer be seen above the instrument noise floor above 4-10 GHz. Separate measurements were carried out using an Agilent E8254A signal generator and an HP8593E spectrum analyzer. Using a 1 kHz resolution bandwidth, the noise floor of the spectrum analyzer at 10 GHz was approximately -100 dBm. Using a source power of 0 dBm, we could determine the frequency above which the filter displays greater than 100 dB of attenuation ($f_{100dB}$). These results, along with $f_{3dB}$, are summarized in table 3.

**Table 3.** Summary of filter bandwidths at different temperatures.

| Dielectric Material | $f_{3dB}$ (GHz), 4.2 K | $f_{3dB}$ (GHz), 77 K | $f_{3dB}$ (GHz), 295 K | $f_{100dB}$ (GHz), 295 K |
|---|---|---|---|---|
| CR-124 | 0.56 | 0.54 | 0.46 | 4.9 |
| FGM-40 | 0.58 | 0.54 | 0.45 | 7.2 |
| GDS | 0.83 | 0.79 | 0.75 | 12.1 |
| MCS | 0.72 | 0.41 | 0.34 | 5.9 |

Modifying the length of the stripline will modify the attenuation, enabling the bandwidth to be optimized for a particular application. In Figure 3, we use the measured attenuation for a length of 32 mm to extrapolate the 3 dB bandwidth as a function of stripline length at both room temperature and 4.2 K. Given the inherent tradeoff between bandwidth and stopband attenuation, we note that filters with bandwidths much greater than 1 GHz may no longer have sufficient high frequency attenuation for some applications.

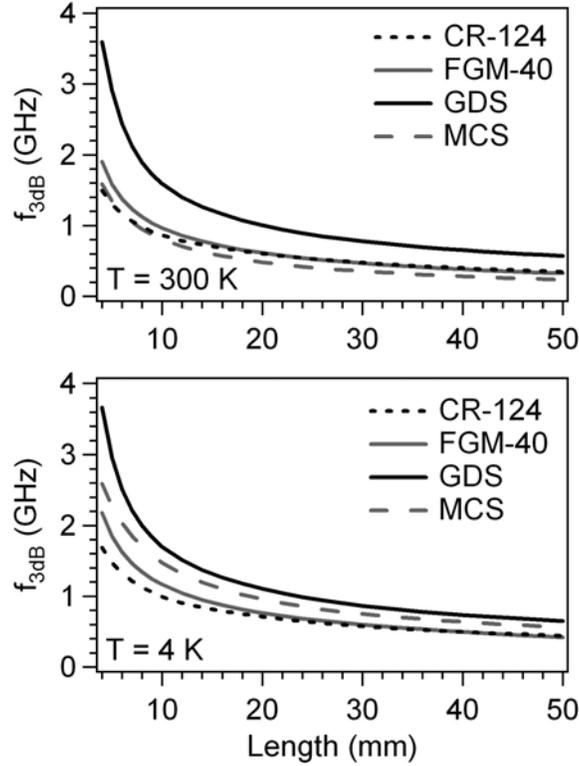

**Figure 3.** Calculated 3 dB bandwidth for filters of different stripline lengths at both room temperature (top) and 4.2 K (bottom).

The return loss is approximately the same for all four filters and is highly sensitive to the coaxial-to-stripline transition. The return loss of the GDS stripline filter is plotted in figure 4. At room temperature, the return loss is greater than 15 dB at all frequencies. The return loss is slightly lower at low temperature, which is likely related to a temperature-dependent permittivity, but remains greater than 10 dB at all temperatures measured. (The width of the center conductor was chosen based on room temperature measurements.) An improved coaxial-to-stripline transition or a filter made in a coaxial geometry could likely achieve a somewhat greater return loss. However, the average return loss over this frequency range is limited by the fact that the magnetic permeability, and hence the impedance, changes with frequency.

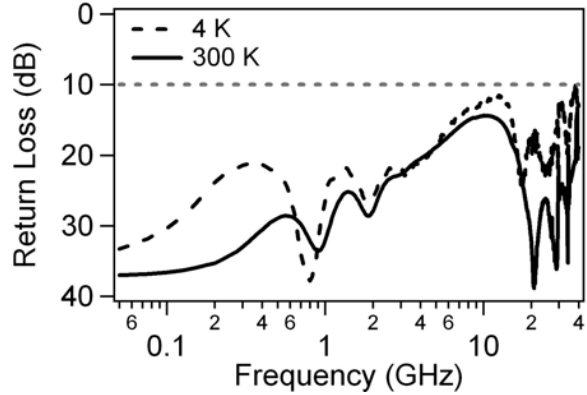

**Figure 4.** Return loss of the GDS stripline filter at room temperature and 4.2 K. For reference, a return loss of 10 dB (10% reflected power) is indicated by the grey dashed line.

A sufficiently large external field will saturate the internal magnetization, decreasing the permeability and hence the attenuation. At room temperature, $f_{3dB}$ of the MCS stripline filter increased from 0.34 to 0.65 GHz with an external field of approximately 0.1 T applied perpendicular to both the direction of signal propagation and the flat side of the center conductor. The field was applied with a permanent magnet, and the field strength was measured with a gaussmeter. The filter bandwidth returned to its original zero-field value upon removal of the external field.

Performance of the filters at sub-Kelvin temperatures has not yet been tested. One potential issue is thermalization. As the filter is strongly absorbing in its stopband, it will also have be an efficient emitter of thermal noise, with a noise power given by equation 1. In the Rayleigh-Jeans limit, this simplifies to the Johnson noise result, with a noise power per unit bandwidth of $k_BT$, where T is the physical temperature of the filter dielectric material. For measurements of ultra-sensitive devices, having the filter at low temperature is important for minimizing the noise power seen by the device. We expect that the CR-124 epoxy material will be easier to heat sink than the silicone material.

**4. Conclusion**

We have constructed and characterized magnetically lossy low-pass stripline filters for cryogenic measurements of highly sensitive non-linear devices. These filters simultaneously achieve a broad, high-attenuation stopband and high return loss measured out to 40 GHz. The 3 dB bandwidth is between 0.3 and 0.8 GHz and can be adjusted beyond this range by changing the filter length. We also note that a significantly sharper roll-off can be achieved with only a modest change in the return loss by following the stripline filter with a conventional reactive filter of an appropriate bandwidth.

**Acknowledgements**

We thank M H Devoret, V Manucharyan and F P Milliken for helpful discussions. This work was supported by NSF-CHE, NSF-DMR and Yale University.

**References**
[1] Schiffres P 1964 A dissipative coaxial RFI filter *IEEE Trans. Electromag. Compat.* **6** 55-61.
[2] Denny H W and Warren W B 1968 Lossy transmission line filters *IEEE Trans. Electromag. Compat.* **10** 363-70.
[3] Vion D, Orfila P F, Joyez P, Esteve D, and Devoret M H 1995 Miniature electrical filters for single-electron devices *J. Appl. Phys.* **77** 2519-24.
[4] Santavicca D F, Reese M O, True A B, Schmuttenmaer C A, Prober D E 2007 Antenna-coupled niobium bolometers for terahertz spectroscopy *IEEE Trans. Appl. Supercond.* **17** 412-5.
[5] Martinis J M, Devoret M H, and Clarke J 1987 Experimental tests for the quantum behavior of a macroscopic degree of freedom: The phase difference across a Josephson junction *Phys. Rev. B* **35** 4682-98.
[6] Fukushima A, Sato A, Iwasa A, Nakamura Y, Komatsuzaki T, and Sakamoto Y 1997 Attenuation of microwave filters for single-electron tunneling experiments *IEEE Trans. Instrum. Meas.* **46** 289-93.
[7] Bladh K, Gunnarsson D, Hurfeld E, Devi S, Kristoffersson C, Smalander B, Pehrson S, Claeson T, Delsing P, and Taslakov M 2003 Comparison of cryogenic filters for use in single electron experiments *Rev. Sci. Instrum.* **74** 1323-7.
[8] Lukashenko A and Ustinov A V 2008 Improved powder filters for qubit measurements *Rev. Sci. Instrum.* **79** 014701.
[9] Jin I, Amar A, and Wellstood F C 1997 Distributed microwave damping filters for superconducting quantum interference devices *Appl. Phys. Lett.* **70** 2186-8.
[10] le Sueur H and Joyez P 2006 Microfabricated electromagnetic filters for millikelvin experiments *Rev. Sci. Instrum.* **77** 115102.
[11] Zorin A B 1995 The thermocoax cable as the microwave frequency filter for single electron circuits *Rev. Sci. Instrum.* **66** 4296-300.
[12] Milliken F P, Rozen J R, Keefe G A, and Koch R H 2007 50 Ω characteristic impedance low-pass metal powder filters *Rev. Sci. Instrum.* **78** 024701.
[13] Park M-J, Choi J, and Kim S-S 2000 Wide bandwidth pyramidal absorbers of granular ferrite and carbonyl iron powders *IEEE Trans. Mag.* **36** 3272-3274.
[14] Emerson and Cuming Microwave Products (Randolph, MA).
[15] Applied Engineering Products (New Haven, CT), part #9308-1113-001.